\documentclass[aps,amsmath,amssymb,twocolumn]{revtex4}

\usepackage{amsmath,amssymb,amsthm,bm,mathtools,mathrsfs,microtype,graphicx,ulem,xcolor}

\usepackage{tikz}
\usetikzlibrary{arrows.meta,positioning,calc}
\usepackage{pgfplots}
\pgfplotsset{compat=1.18}
\usepackage{xcolor}
\usepgfplotslibrary{groupplots}
\usepackage{capt-of}
\usepackage[hypertexnames=false]{hyperref}

\begin{document}

\title{Optimal quantum discrimination at Liouvillian exceptional points}
\author{Aritra Ghosh\footnote{aritraghosh500@gmail.com} and M. Bhattacharya}
\affiliation{School of Physics and Astronomy, Rochester Institute of Technology, 84 Lomb Memorial Drive, Rochester, New York 14623, USA}
\vskip-2.8cm
\date{\today}
\vskip-0.9cm

\vspace{5mm}
\begin{abstract}
Central to non-Hermitian sensing is the question of whether exceptional points (EPs) can make weak perturbations easier to detect. We show that despite enhanced spectral response, in finite-dimensional Lindblad systems with a unique stationary state and a nonzero relaxation gap, EPs offer no fundamental advantage in how the observation time required for reliable detection of a weak perturbation scales with the perturbation strength. We nevertheless demonstrate that by selectively choosing perturbations that enhance the quantum Fisher information rate, EPs can shorten the observation time needed to reach a predefined detection accuracy relative to a comparable non-EP reference at the same decay rate. As an illustration, we show using an optomechanical model at an EP that we can detect a weak force through the cavity output at the same reliability in less time than a non-EP reference. Our results establish fundamental limits on perturbation detection at EPs and can guide the design of faster non-Hermitian sensors based on optomechanical resonators, superconducting qubits, and trapped ions.
\end{abstract}
\maketitle

\begin{figure*}[t!]
\centering

\definecolor{epnavy}{HTML}{17324D}
\definecolor{epteal}{HTML}{148C86}
\definecolor{epcoral}{HTML}{DB5F54}
\definecolor{epgold}{HTML}{C8941A}
\definecolor{epgray}{HTML}{697A89}

\resizebox{0.96\textwidth}{!}{%
\begin{tikzpicture}[
    font=\sffamily,
    >={Latex[length=2.2mm,width=1.3mm]},
    tick/.style={
        line width=.75pt
    },
    result/.style={
        font=\bfseries\small
    }
]

\node[
    rounded corners=2.5pt,
    draw=epnavy!24,
    fill=epnavy!3,
    line width=.7pt,
    minimum width=3.75cm,
    minimum height=2.30cm
] (spec) at (-4.75,0) {};

\node[
    font=\bfseries\small,
    text=epnavy
] at ([yshift=-.29cm]spec.north)
    {finite relaxation time};

\begin{scope}[shift={(-4.75,-.10)}]

\draw[
    ->,
    line width=.7pt,
    color=epgray
] (-1.42,0)--(1.43,0);

\fill[epcoral] (-.72,0) circle (2.7pt);

\draw[
    epcoral,
    line width=.85pt
] (-.72,0) circle (5.2pt);

\node[
    font=\scriptsize,
    text=epcoral!85!black,
    above=4pt
] at (-.72,0) {decay EP$m$};

\fill[epteal] (1.02,0) circle (2.7pt);

\node[
    font=\scriptsize,
    text=epteal!75!black,
    above=4pt
] at (1.02,0) {steady state};

\draw[
    <->,
    line width=.7pt,
    color=epgold!90!black
] (-.54,-.47)--(.84,-.47)
node[
    midway,
    below=1pt,
    font=\scriptsize,
    text=epgold!75!black
] {$\gamma_{\rm EP}>0$};

\end{scope}

\draw[
    epnavy!15,
    line width=.7pt
] (-2.48,-1.14)--(-2.48,1.34);

\draw[
    ->,
    line width=1.05pt,
    color=epnavy!52
] (-1.80,0)--(6.05,0);

\path[
    fill=epcoral!10
]
(-1.68,.02)
.. controls (-1.30,.76) and (-.58,.78) .. (-.18,.43)
.. controls (.28,.16) and (.72,.06) .. (1.22,.02)
-- cycle;

\draw[
    epcoral!82,
    line width=1.15pt
]
(-1.68,.02)
.. controls (-1.30,.76) and (-.58,.78) .. (-.18,.43)
.. controls (.28,.16) and (.72,.06) .. (1.22,.02);

\node[
    font=\scriptsize\bfseries,
    text=epcoral!82!black
] at (-.66,1.08) {EP$m$ relaxation};

\draw[
    tick,
    color=epgold!90!black
] (-.62,-.11)--(-.62,.11);

\node[
    anchor=north,
    font=\scriptsize,
    text=epgold!75!black
] at (-.62,-.16)
    {$T_{\rm EP}\sim\gamma_{\rm EP}^{-1}$};

\fill[white] (1.82,0) circle (4.1pt);

\draw[
    epcoral,
    line width=1.25pt
] (1.82,0) circle (4.1pt);

\node[
    anchor=south,
    font=\scriptsize\bfseries,
    text=epcoral!82!black
] at (1.82,.18)
    {splitting time};

\node[
    anchor=north,
    font=\scriptsize
] at (1.82,-.18)
    {$T_{\rm split}\sim|\delta|^{-1/m}$};

\node[
    anchor=north,
    result,
    text=epcoral!82!black
] at (1.82,-.57)
    {$P\to50\%$};

\fill[white] (5.05,0) circle (4.1pt);

\draw[
    epteal,
    line width=1.25pt
] (5.05,0) circle (4.1pt);

\node[
    anchor=south,
    font=\scriptsize\bfseries,
    text=epteal!75!black
] at (5.05,.18)
    {detection time};

\node[
    anchor=north,
    font=\scriptsize
] at (5.05,-.18)
    {$T_{\rm disc}\gtrsim|\delta|^{-2}$};

\node[
    anchor=north,
    result,
    text=epteal!75!black
] at (5.05,-.57)
    {e.g. $90\%$ correct};

\draw[
    <->,
    line width=.75pt,
    color=epnavy!55
] (2.14,.76)--(4.73,.76)
node[
    midway,
    fill=white,
    inner sep=2pt,
    font=\scriptsize\bfseries,
    text=epnavy
] {$T_{\rm disc}/T_{\rm split}\to\infty$};

\end{tikzpicture}
}

\caption{\textbf{Separation of timescales.} For the finite-dimensional gapped Lindblad systems, a generic perturbation $\delta$ of an EP of order $m$, the best probability $P$ of correctly deciding whether it is present approaches $50\%$ on the inverse-splitting timescale as $\delta\to0$. Presence and absence are equally likely. Reliable detection requires much longer observation times.}
\label{fig1}

\end{figure*}
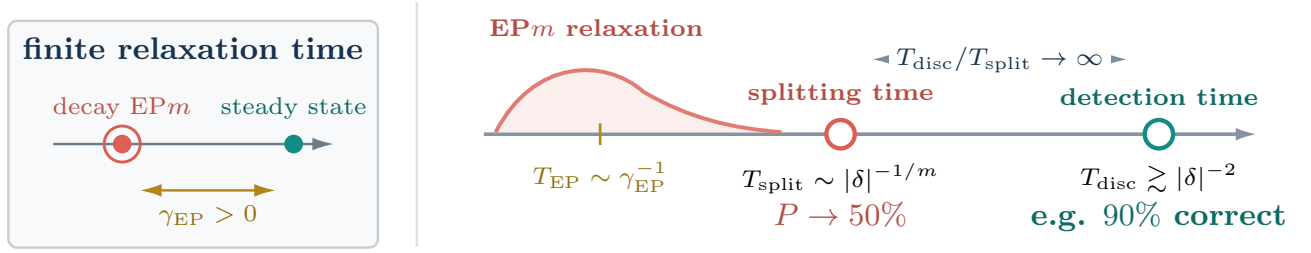

\section{Introduction} 
An exceptional point (EP) is a non-Hermitian degeneracy at which both eigenvalues and eigenvectors coalesce \cite{Kato_1995,Berry_2004,Heiss_2012,Miri_2019,Ashida_2020,Bergholtz_2021}. A generic perturbation $\delta$ of an EP of order $m$ (EP$m$) separates the coalesced eigenvalues by $\sim|\delta|^{1/m}$ \cite{Kato_1995,Ashida_2020,Wiersig_2022a,Wiersig_2022b} and sensing using EPs has attracted immense interest because of this enhanced spectral response
\cite{Wiersig_2014,Wiersig_2016,Chen_2017,Hodaei_2017,Chen_2018,Lai_2019,Hokmabadi_2019,Park_2020,Wiersig_2020}. While existing work has established precision, signal-to-noise ratios, and Fisher-information bounds, whether or not EPs can provide a fundamental advantage for quantum sensing remains debated
\cite{Langbein_2018,Lau_2018,Zhang_2019,Chen_2019,Ding_2023,Loughlin_2024,Peng_2024}. These limits still allow useful finite improvements
\cite{Wiersig_2026}, while squeezing can supply an
additional resource \cite{Wang_2026}.

A practical test of an EP-based sensing advantage is whether it reduces the observation time needed to decide reliably whether a weak perturbation is present or not. Consider a specified weak perturbation $\delta$ that is either absent or present, with the two cases equally likely. The task is to determine from quantum measurements which case applies, which is a primary goal of quantum discrimination  \cite{Holevo_1973,Helstrom_1976,Barnett_2009,Watrous_2018,Molmer_2015,Kiilerich_2018}. Notably, even random guessing gives a $50\%$ probability of a correct answer. Thus reliable detection means reaching a prescribed probability of finding the correct answer, such as $90\%$, for a weak perturbation. The eigenvalue-splitting scale of an EP describes through its inverse the natural timescale over which the split modes begin to appreciably differ, say, via the accumulation of phase factors. This timescale, however, may not be sufficient for an observation to make the perturbation detectable with reliable accuracy. More broadly, we ask whether operating an open quantum system at its EP can reliably detect a weak perturbation in less time than a comparable reference sensor without an EP, while maintaining the same detection accuracy.

In this work we provide, to our knowledge, the first
quantum-discrimination benchmark for weak-perturbation sensing at Liouvillian EPs and an exact balance governing the detection advantages available for different Hamiltonian perturbations. Related studies have distinguished prepared states using non-Hermitian evolution \cite{Chen_QSD_2022},
optimized minimum-error target detection in optical
scattering \cite{Bouchet_2021}, and quantified quantum Fisher information (QFI) enhancement at EPs within a scattering framework \cite{Wiersig_2026}. We pose EP sensing as a quantum-discrimination task \cite{Holevo_1973,Helstrom_1976,Barnett_2009,Watrous_2018} and determine how Liouvillian EPs affect the observation time and maximum probability of detecting a weak perturbation by optimally distinguishing the quantum states generated in its presence and absence. By optimal, we mean that arbitrary quantum measurements are allowed on all signals emitted by the sensor during the observation together with the sensor's state at the end of the observation window  \cite{Molmer_2015,Kiilerich_2018}. This includes every measurement of a smaller portion of the signal, since unwanted signals can always be ignored. A fundamental limit with complete access therefore also constrains measurements with more limited access \cite{Watrous_2018}.

Our general analysis concerns finite-dimensional Markovian open systems governed by a Lindblad generator \cite{Gorini_1976,Lindblad_1976}, with a unique stationary state and a nonzero relaxation gap. Such systems are pervasive in the quantum technologies, including superconducting qubits \cite{Chen_2022}, photonic qubits coupled to superconducting quantum circuits \cite{Hu_2018}, trapped ions \cite{Wu_2026}, and qubits implemented on quantum processors \cite{Abo_2024}, and have been experimentally demonstrated to admit EPs \cite{Chen_2022,Wu_2026,Abo_2024}. The gap keeps relaxation times finite as the EP is approached and an EP among the decaying modes produces relaxation dynamics with powers of time multiplying an exponential \cite{Ashida_2020,Minganti_2019,Minganti_2020,Khandelwal_2021}. Within our broad assumptions, our general results apply across sensor platforms and do not rely on a particular Hamiltonian or environmental coupling.

We first show that for generic perturbations of the EP, even the best quantum measurement approaches accuracy no better than randomly guessing the outcome on the timescale dictated by the EP splitting. A fixed probability of a correct answer better than random guessing requires an observation time growing at least as the inverse square of the perturbation strength, consistent with the results of quantum Markov theory \cite{Catana_2015,Guta_2017}.
We then establish an exact information balance for
a class of Hamiltonian perturbations associated with
the coalescing decay modes. As we show, compared with a non-EP reference baseline, an EP can shorten the time needed to detect one perturbation while increasing the time needed to detect another perturbation. We identify which perturbations benefit and quantify how much faster they can be detected at the same probability of obtaining the correct answer as the non-EP reference, providing a guide for choosing EP sensors for specific sensing tasks. We demonstrate this selective advantage for force sensing \cite{Tsang_2012,Mason_2019} in an optomechanical model \cite{Aspelmeyer_2014,Ghosh_2026}, which is an infinite-dimensional example suggesting that our conclusions can extend beyond the finite-dimensional systems covered by our general proofs. By choosing a weak mechanical force as the perturbation to detect, we show that optimal measurements of the ideally collected cavity output require approximately $49.85\%$ less observation time than an ideally observed non-EP reference oscillator to reach the same detection reliability.

\section{Spectral splitting to detection}
We consider a finite-dimensional open quantum system whose density matrix evolves according to the master equation $\dot{\rho}=\mathcal{L}_\delta(\rho)$, where $\delta=0$ denotes the EP and the Liouvillian is a Lindblad generator \cite{Gorini_1976,Lindblad_1976}
\begin{equation}
\mathcal{L}_\delta(\rho)
=-i[H_\delta,\rho]
+\sum_k L_{k,\delta}\rho L_{k,\delta}^\dagger-\frac12\sum_k
\{L_{k,\delta}^\dagger L_{k,\delta},\rho\},
\label{lindblad-detection}
\end{equation}
where we have taken $\hbar=1$. Each $L_{k,\delta}$ describes coupling to an environmental channel whose input is initially in vacuum, while the index $k$ identifies the same physical channel with and without the perturbation. We assume that $H_\delta$ and $L_{k,\delta}$ are analytic in a neighborhood of $\delta=0$. Throughout this neighborhood, the stationary eigenvalue zero has multiplicity one and every other eigenvalue obeys $\operatorname{Re}[\lambda]\leq-\Delta<0$, with a fixed relaxation gap $\Delta>0$ \cite{Chen_2022,Abo_2024}.

Now to perform the quantum-discrimination task, we prepare the sensor in the same fixed pure state under both the possibilities, i.e., $\delta \neq 0$ and $\delta = 0$. The sensor and the externally emitted signals together evolve unitarily \cite{Catana_2015}, so their combined state remains pure even though the sensor taken by itself can become mixed. We write this combined state at time $T$ as $|\Psi_\delta^{(T)}\rangle$, including all signals emitted during the time period $[0,T]$. Explicitly,
$|\Psi_\delta^{(T)}\rangle=U_\delta(T)(|\psi_{\rm in}\rangle_{\rm sensor}\otimes|\mathrm{vac}\rangle_{\rm inputs})$,
where $U_\delta(T)$ is the joint unitary evolution.
The initial sensor state $|\psi_{\rm in}\rangle$ is arbitrary but identical under both the $\delta=0$ and $\delta \neq 0$ cases that need to be discriminated. Assuming equally likely the presence and absence of the perturbation $\delta$, the largest probability of a correct detection conclusion over all quantum measurements follows from Helstrom's formula \cite{Holevo_1973,Helstrom_1976}
\begin{equation}
P^\star(\delta,T)=\frac{1+\sqrt{1-|A_T(\delta)|^2}}{2},
\label{correct-overlap}
\end{equation}
where $A_T(\delta)=\langle\Psi_0^{(T)}|\Psi_\delta^{(T)}\rangle$ is the overlap between the states of the sensor and its emitted signals without and with the perturbation. Identical states give $P^\star=1/2$, while orthogonal states allow a correct answer with certainty. Measurements using fewer output signals cannot do better, since the allowed measurements can always ignore unwanted signals \cite{Molmer_2015,Kiilerich_2018,Watrous_2018}. For a desired target probability $p\in(1/2,1)$, the shortest observation time is $T_{\rm disc}(p,\delta)=\inf\{T:P^\star(\delta,T)\geq p\}$, with infinity assigned if the target is never reached.

For measurement outcomes $x$ with probabilities
$p_x(\delta)$, the Fisher information is
$F(\delta)=\sum_x[\partial_\delta p_x(\delta)]^2/p_x(\delta)$, with continuous outcomes requiring an integral replacing the sum. It quantifies how sensitively the outcome probabilities change with the perturbation. Maximizing over quantum measurements gives the QFI \cite{Braunstein_1994}, with $\mathcal{I}_Q(T,0)$ denoting its value at the EP. Its long-time growth is $\mathcal{I}_Q(T,0)=Tf_Q+O(1)$
\cite{Gammelmark_2014,Catana_2015,Guta_2017},
where $f_Q$ is the growth rate of the QFI and $O(1)$
remains bounded as $T$ grows. Related rates are known for Gaussian bosonic sensors \cite{Yokomizo_2026}.

To this end we form $X_{\delta,T}=\operatorname{Tr}_{\rm outputs}\left[|\Psi_\delta^{(T)}\rangle\langle\Psi_0^{(T)}|\right]$, which traces over the emitted signals. This leaves an operator acting only on the sensor, whose trace over the sensor gives the overlap $A_T(\delta)=\operatorname{Tr}[X_{\delta,T}]$. Throughout, $\operatorname{Tr}$ without a subscript denotes the trace over the sensor, whereas $\operatorname{Tr}_{\rm outputs}$ denotes the partial trace over the emitted signals. Now the operator $X_{\delta,T}$ obeys $\dot{X}_{\delta,T}=M_\delta(X_{\delta,T})$,
where $M_\delta$ contains the perturbed Hamiltonian and couplings acting on the left and their unperturbed counterparts acting on the right, so that at $\delta=0$, $M_0=\mathcal{L}_0$.
The simple zero eigenvalue of $M_0$, denoted by $\chi(0)=0$, continues for $\delta\neq0$ to an analytic eigenvalue $\chi(\delta)$, while the gap ensures that this eigenvalue controls the long-time overlap $A_T\sim\alpha(\delta)e^{T\chi(\delta)}$, where the prefactor $\alpha(\delta)$ is analytic with $\alpha(0)=1$. The bound $|A_T|\leq1$ excludes a linear term in $\operatorname{Re}[\chi(\delta)]$, while the pure-state QFI determines its quadratic term $\operatorname{Re}[\chi(\delta)]=-f_Q\delta^2/8+O(\delta^3)$, with finite $f_Q\geq0$ (End Matter). Substituting this overlap into Eq. (\ref{correct-overlap}) gives the following detection limit:
\begin{equation}
P^\star(\delta,T)
=\frac{1+\sqrt{1-e^{-f_Q\delta^2T/4}}}{2}+o(1).
\label{local-correct}
\end{equation}
Here $\delta\to0$ and $T\to\infty$ while $\delta^2T$
remains bounded, and $o(1)$ vanishes in this limit.

Let $\lambda_i(\delta)$ be the eigenvalues emerging from the decay EP, so that their largest separation scales as $\Omega(\delta)=\max_{i,j}|\lambda_i(\delta)-\lambda_j(\delta)|$. A mode evolves as $e^{\lambda_iT}$, so differences between
eigenvalues multiplied by time describe differences in decay exponents or accumulated phases. Writing $\Omega(\delta)\sim c_\Omega|\delta|^\beta$ with
$c_\Omega>0$, we define the associated timescale as $T_{\rm split}=\tau/\Omega(\delta)$ for fixed $\tau>0$. It then follows that for $0<\beta<2$, the limit $\delta \rightarrow 0$ gives $\delta^2T_{\rm split}=O(|\delta|^{2-\beta})\longrightarrow0$, substituting which into Eq. (\ref{local-correct}) directly gives $P^\star(\delta,T_{\rm split})\longrightarrow 1/2$. Generic perturbations of an EP of order $m = 2,3,\cdots$ have $\beta=1/m$, which is well within $0<\beta<2$. Thus the inverse-splitting timescale is insufficient for reliable detection, as the probability of determining whether a perturbation $\delta \rightarrow 0$ exists or not via the best quantum measurements is no better than randomly guessing its existence. Thus, while EP transients can remain visible during relaxation \cite{Minganti_2019,Khandelwal_2021}, detecting a progressively weaker perturbation requires much longer observations. We focus on perturbations for which $f_Q>0$, so any fixed target probability $p\in(1/2,1)$ is reached on the inverse-square timescale $\delta^2T_{\rm disc}(p,\delta)\to 4\log[1/(4p(1-p))]/f_Q$ from Eq. (\ref{local-correct}). We illustrate this timescale separation in Fig. \ref{fig1}. However, there is an alternate route in which EPs can produce finite yet nonzero detection-time advantages through certain selective Hamiltonian perturbations, as we show next.

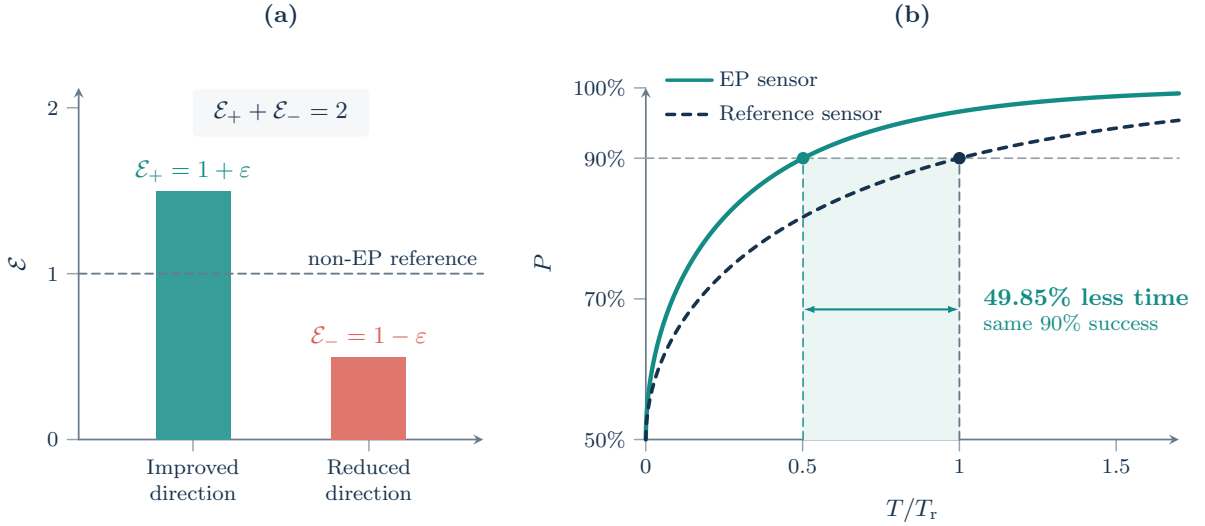
\begin{figure*}[t]
\centering
\begingroup

\definecolor{gainNavy}{HTML}{17324D}
\definecolor{gainTeal}{HTML}{148C86}
\definecolor{gainCoral}{HTML}{DB5F54}
\definecolor{gainGray}{HTML}{697A89}

\pgfmathsetmacro{\epGain}{1.99400935}
\pgfmathsetmacro{\epTime}{1/\epGain}

\begin{tikzpicture}[
    font=\small,
    >={Latex[length=1.5mm,width=1.0mm]},
    line cap=round,
    line join=round
]

\pgfplotsset{
    gainAxis/.style={
        scale only axis,
        axis lines=left,
        axis line style={
            draw=gainNavy!65,
            line width=.65pt
        },
        tick align=outside,
        tick style={
            draw=gainNavy!45,
            line width=.55pt
        },
        ticklabel style={
            font=\footnotesize,
            text=gainNavy
        },
        label style={
            font=\small,
            text=gainNavy
        },
        title style={
            font=\bfseries\small,
            text=gainNavy
        },
        clip=false,
        every axis plot/.append style={
            line cap=round
        }
    }
}

\begin{axis}[
    gainAxis,
    at={(0,0)},
    anchor=south west,
    width=5.35cm,
    height=4.65cm,
    xmin=.35,
    xmax=2.65,
    ymin=0,
    ymax=2.12,
    xtick={1,2},
    xticklabels={
        {Improved\\direction},
        {Reduced\\direction}
    },
    x tick label style={
        font=\footnotesize,
        align=center
    },
    ytick={0,1,2},
    ylabel={$\mathcal{E}$},
    ylabel style={yshift=2pt},
    title={(a)},
    title style={
        at={(.5,1.10)},
        anchor=south
    }
]

\addplot[
    ybar,
    bar width=28pt,
    draw=none,
    fill=gainTeal!85
]
    coordinates {(1,1.5)};

\addplot[
    ybar,
    bar width=28pt,
    draw=none,
    fill=gainCoral!85
]
    coordinates {(2,.5)};

\draw[
    gainNavy!65,
    densely dashed,
    line width=.7pt
]
    (axis cs:.35,1)--(axis cs:2.65,1);

\node[
    anchor=south east,
    font=\footnotesize,
    text=gainNavy,
    inner sep=0pt,
    yshift=3pt
]
    at (axis cs:2.62,1)
    {non-EP reference};

\node[
    anchor=south,
    text=gainTeal,
    font=\small,
    inner sep=3pt
]
    at (axis cs:1,1.5)
    {$\mathcal{E}_+=1+\varepsilon$};

\node[
    anchor=south,
    text=gainCoral,
    font=\small,
    inner sep=3pt
]
    at (axis cs:2,.5)
    {$\mathcal{E}_-=1-\varepsilon$};

\node[
    text=gainNavy,
    font=\small,
    fill=gainNavy!4,
    rounded corners=2pt,
    inner xsep=8pt,
    inner ysep=5pt
]
    at (axis cs:1.5,1.97)
    {$\mathcal{E}_++\mathcal{E}_-=2$};

\end{axis}

\begin{axis}[
    gainAxis,
    name=probability,
    at={(7.50cm,0)},
    anchor=south west,
    width=7.05cm,
    height=4.65cm,
    xmin=0,
    xmax=1.70,
    ymin=.50,
    ymax=1.00,
    xtick={0,.5,1,1.5},
    xticklabels={$0$,$0.5$,$1$,$1.5$},
    ytick={.5,.7,.9,1},
    yticklabels={$50\%$,$70\%$,$90\%$,$100\%$},
    xlabel={$T/T_{\rm r}$},
    xlabel style={yshift=-3pt},
    ylabel={$P$},
    ylabel style={yshift=2pt},
    title={(b)},
    title style={
        at={(.5,1.10)},
        anchor=south
    },
    legend style={
        at={(.035,1.00)},
        anchor=north west,
        yshift=3mm,
        draw=none,
        fill=none,
        font=\footnotesize,
        text=gainNavy,
        cells={anchor=west},
        row sep=2pt,
        inner sep=0pt
    },
    legend image post style={
        line width=1.3pt
    }
]

\path[fill=gainTeal!9]
    (axis cs:\epTime,.50)
    rectangle
    (axis cs:1,.90);

\draw[
    gainGray!70,
    densely dashed,
    line width=.65pt
]
    (axis cs:0,.90)--(axis cs:1.70,.90);

\draw[
    gainTeal!80,
    densely dashed,
    line width=.7pt
]
    (axis cs:\epTime,.50)--(axis cs:\epTime,.90);

\draw[
    gainNavy!65,
    densely dashed,
    line width=.7pt
]
    (axis cs:1,.50)--(axis cs:1,.90);

\addplot[
    gainTeal,
    line width=1.65pt,
    domain=0:1.303840481,
    samples=251,
    variable=\u
]
    ({\u^2},
     {.5*(1+sqrt(max(0,
         1-exp(ln(.36)*\epGain*\u^2)
     )))});
\addlegendentry{EP sensor}

\addplot[
    gainNavy,
    line width=1.35pt,
    dashed,
    domain=0:1.303840481,
    samples=251,
    variable=\u
]
    ({\u^2},
     {.5*(1+sqrt(max(0,
         1-exp(ln(.36)*\u^2)
     )))});
\addlegendentry{Reference sensor}

\fill[gainTeal]
    (axis cs:\epTime,.90)
    circle[radius=2.35pt];

\fill[gainNavy]
    (axis cs:1,.90)
    circle[radius=2.35pt];

\draw[
    <->,
    gainTeal,
    line width=.8pt
]
    (axis cs:\epTime,.685)--(axis cs:1,.685);

\node[
    anchor=west,
    align=left,
    text=gainTeal,
    inner sep=0pt
]
    at (axis cs:1.075,.685)
    {
        \textbf{49.85\% less time}\\[-1pt]
        {\footnotesize same $90\%$ success}
    };

\end{axis}

\end{tikzpicture}
\endgroup

\caption{
\textbf{Finite improvements in EP-based detection.}
(a) For a single real EP2 decay block, the most- and least-informative perturbation directions have relative information rates $\mathcal{E}_\pm=1\pm \varepsilon$, with $0<\varepsilon\leq1$; the illustration uses $\varepsilon=1/2$. (b) For a selected weak mechanical force and ideal collection of the cavity emission ($\eta=1$), the optomechanical EP sensor reaches a $90\%$ probability of a correct decision with $49.85\%$ less observation time than a non-EP reference oscillator with the same decay rate. The curves show the weak-force limit and the time is normalized to the reference oscillator's time to reach the same $90\%$ target.}
\label{fig2}
\end{figure*}

\section{Finite detection gains at an exceptional point}
The preceding detection limit makes the information rate $f_Q$ central to comparing sensors. We now ask how the relaxation dynamics at an EP affect this rate for Hamiltonian perturbations, complementing studies of QFI within scattering formalisms
\cite{Wiersig_2026,Horodynski_2021}. We begin with the perturbed Hamiltonian $H_\delta=H_0+\delta V$, with Hermitian $V$ and fixed environmental couplings, i.e., jump operators. Denoting by $\pi$ the unique stationary state satisfying $\mathcal{L}_0(\pi)=0$, we take $\operatorname{Tr}(\pi V)=0$ and define $\sigma_V^2=\operatorname{Tr}(\pi V^2)$. With the time-evolving operator $V(t)=e^{t\mathcal{L}_0^\dagger}V$, one can write \cite{Catana_2015,Guta_2017}
\begin{equation}
\frac{f_Q(V)}{8}
=\int_0^\infty
\frac12\operatorname{Tr}\!\left[\pi\{V,V(t)\}\right]dt,
\label{correlation-rate}
\end{equation}
whose integrand we denote by $C_V(t)$, which measures how the fluctuations of $V$ with respect to the stationary state $\pi$ remain correlated as $V$ evolves in time, with $C_V(0)=\sigma_V^2$. For a fixed variance $\sigma_V^2$, the $f_Q(V)$ corresponding to a perturbing operator $V$ is therefore determined by the area under the correlation curve in Eq. (\ref{correlation-rate}), including any negative contributions. We focus on one EP$m$ of $\mathcal{L}_0^\dagger$ at the real decay eigenvalue $-\gamma_{\rm EP}<0$. Its relaxation requires $m$ independent Hermitian operators whose time dependence contains polynomial factors multiplying $e^{-\gamma_{\rm EP}t}$ \cite{Ashida_2020}. We restrict $V$ to their real linear combinations, so the relaxation dynamics keeps the evolving operator $V(t)$ within this space. We further assume that every nonzero real linear combination of these $m$ observables has a positive stationary variance.

For comparison, we introduce a non-EP sensor as a reference, whose correlations decay as a simple exponential. Matching its unperturbed stationary state, variance, and exponential decay rate to those of the EP sensor gives $C_{\rm r}(t)=\sigma_V^2e^{-\gamma_{\rm EP}t}$ and hence $f_{\rm r}=8\sigma_V^2/\gamma_{\rm EP}$ from Eq. (\ref{correlation-rate}). This benchmark provides a direct comparison, i.e., the correlation curves start at the same value and share the same exponential decay but the EP introduces additional polynomial contributions via the Jordan dynamics which the non-EP reference lacks. We therefore evaluate how these polynomial contributions change the information rate through Eq. (\ref{correlation-rate}). We compare different choices for perturbing operators $V$ formed from real linear combinations of the $m$ Hermitian observables describing the coalescing decay modes, with each $V$ admitting the stationary variance $\operatorname{Tr}[\pi V^2]=\sigma^2$. Their correlations take the form $C_V(t)=e^{-\gamma_{\rm EP}t}[\sigma^2+\sum_{k=1}^{m-1}c_k(V)t^k]$, with coefficients depending on the chosen perturbation. Substituting this expression into Eq. (\ref{correlation-rate}), the constant term reproduces the reference rate $f_{\rm r}$, while the polynomial terms provide the change due to EP relaxation. Their integrated contribution increases the information rate for some perturbations and decreases it for others. Averaging uniformly over all these $V$, the polynomial contributions cancel exactly, giving $\overline{f_Q}=f_{\rm r}$, showing no averaged advantage when considering all different possible perturbations. However, a tantalizing selective advantage remains. Considering a particular perturbation $V$, the quadratic dependence of $f_Q$ on $V$, together with its nonnegativity, however yields $f_Q(V)\leq mf_{\rm r}$ (End Matter). So defining the relative information gain $\mathcal{E}(V)=f_Q(V)/f_{\rm r}$, we obtain
\begin{equation}
\overline{\mathcal{E}}=1,
\qquad
0\leq\mathcal{E}(V)\leq m .
\label{EP-balance}
\end{equation}
Importantly, the information rates for different perturbations differ as we can show by examining the constant and linear terms of an EP's polynomial transient. We choose two observables $W_1,W_2$ whose exact relaxation has the form $W_1(t)=e^{-\gamma_{\rm EP}t}W_1$ and $W_2(t)=e^{-\gamma_{\rm EP}t}(W_2+gtW_1)$, where $g>0$. The pair can be chosen with $\operatorname{Tr}(\pi W_1^2)=\operatorname{Tr}(\pi W_2^2)=1$ and $\operatorname{Tr}[\pi\{W_1,W_2\}]/2=0$. Constructing now two perturbing operators $V_\pm=\sigma(W_1\pm W_2)/\sqrt{2}$, both
of them have stationary variance $\sigma^2$ and their correlations are $C_\pm(t)=\sigma^2e^{-\gamma_{\rm EP}t}(1\pm gt/2)$. These additional linear-time terms have opposite signs and nonzero integrals in Eq. (\ref{correlation-rate}), establishing $\mathcal{E}(V_+)>1$ and $\mathcal{E}(V_-)<1$ [Fig.~\ref{fig2}(a)]. Hence some perturbations necessarily provide more information than the non-EP reference, while others provide less. Since for a fixed target probability $p\in(1/2,1)$, Eq. (\ref{local-correct}) requires the same leading value of $f_Q\delta^2T$ for both the EP and non-EP sensors, their required observation times are inversely proportional to their information rates. For $\mathcal{E}(V)>0$ and an EP$m$, one has from Eq. (\ref{EP-balance}), the ratio
\begin{equation}
\frac{T_{\rm disc}^{\rm EP}(p,\delta)}
     {T_{\rm disc}^{\rm r}(p,\delta)}
\longrightarrow\frac{1}{\mathcal{E}(V)} \geq \frac{1}{m}
\qquad(\delta\to0).
\label{gain-detection}
\end{equation}
An enhancement $\mathcal{E}(V)>1$ therefore shortens the time needed to reach the same detection reliability. Eq. (\ref{gain-detection}) provides a fundamental bound to maximum EP-induced advantage that can be obtained. The lower bound in Eq. (\ref{gain-detection}) is saturated when the information rate is concentrated entirely into the selected perturbation $V$ and higher-order EPs can potentially produce stronger advantages; an EP10 can therefore lead to a maximum advantage by an order-of-magnitude. This, however, should not be interpreted as a universal advantage between EP and non-EP sensors, since our comparison is between the EP sensor and the non-EP reference with equal decay rate and stationary statistics.

\section{Optomechanical force detection}
We now illustrate the selective detection advantage with an optomechanical model, where a weak mechanical force changes the light leaving the cavity. We calculate the output states directly using the linearized cavity optomechanics, where the rotating-wave approximation at the red-sideband resonance gives the following interaction Hamiltonian \cite{Aspelmeyer_2014}: 
\begin{equation}
H_\epsilon=G(a^\dagger b+ab^\dagger)+\epsilon(b+b^\dagger),
\label{opto-model}
\end{equation}
where $a$ and $b$ are the annihilation operators of the cavity and mechanical excitations, respectively, $G$ is the linearized optomechanical coupling, and $\epsilon$ denotes the amplitude of a resonant mechanical force of known phase, applied from $t=0$. We assume that $\kappa_c>\gamma_m>0$ and $G\neq0$ throughout the observation and that both the optical and mechanical modes dissipate into baths with the respective jump operators $L_c=\sqrt{\kappa_c}a$ and $L_m=\sqrt{\gamma_m}b$, the latter motivated by cryogenic-resonator experiments \cite{Satzinger_2018,Omahen_2026}.

Unlike the generic perturbations $\delta$ discussed in the spectral-splitting arguments, the force $\epsilon$ does not split the EP. It changes the mean amplitudes while preserving the decay eigenvalues and their coalescence, as can be checked through a deduction of the adjoint Liouvillian evolution $\mathcal{L}^\dagger_\epsilon (a)$ and $\mathcal{L}^\dagger_\epsilon (b)$. Thus the EP in the two-mode dynamics is the EP2 as for $\epsilon=0$, studied in \cite{Ghosh_2026}, which occurs at $G = G_{\rm EP}=(\kappa_c-\gamma_m)/4$. Because for an EP$m$, Eq. (\ref{gain-detection}) gives a maximum advantage by a factor of $m$, here with an EP2, we expect to reach a factor-of-two advantage in the optimal setting. For later use, we define $\Gamma=(\kappa_c+\gamma_m)/4$ and $u=G_{\rm EP}/\Gamma$.

To understand why the choice of perturbation matters, let us compare the mechanical force $\epsilon V_b$, with $V_b=b+b^\dagger$, with a separate case of a weak optical drive $\epsilon V_a$, with $V_a=i(a^\dagger-a)$, at fixed $G$. Both the operators have variance one in the unperturbed vacuum, so the two perturbations are compared at the same amplitude $\epsilon$ and the same stationary variance. At the EP, their stationary correlations can be evaluated to be $C_{V_b}(t)=(1+G_{\rm EP}t)e^{-\Gamma t}$ and $C_{V_a}(t)=(1-G_{\rm EP}t)e^{-\Gamma t}$. The important point is that the extra time-dependent term increases the correlation integral determining the QFI rate in Eq. (\ref{correlation-rate}) for the mechanical force and decreases it for the optical drive. 

Since the mechanical perturbation enhances the QFI rate, we select the system to be a sensor for the mechanical perturbation and ask how much of its advantage survives in measurements of the cavity output. Suppose that only a fraction $0<\eta\leq1$ of the cavity emission is collected. Starting with the cavity and mechanical modes in their joint vacuum and both the environmental fields in vacuum, the linear dynamics produces a product of coherent states for the modes and their outputs. The collected light is therefore not entangled with the mechanics or the uncollected light, so discarding them leaves it in a pure coherent state. For equally likely force absence and presence, the optimal probability of a correct decision is given by the Helstrom formula as
\begin{equation}
P_{\rm opt}^\star(\epsilon,T)
=\frac{1+\sqrt{1-e^{-Q_T}}}{2},
\label{opto-correct}
\end{equation}
where $\langle a\rangle_\epsilon(t)$ is the mean cavity amplitude and $Q_T=\eta\kappa_c\int_0^T
|\langle a\rangle_\epsilon(t)-\langle a\rangle_0(t)|^2dt$ is the mean photon number associated with the force-induced displacement of the collected light. Here $\langle a\rangle_0(t)$ is the cavity amplitude without the force which vanishes for an initial joint vacuum of the
fluctuation modes. The cavity amplitude approaches $\langle a\rangle_\epsilon(\infty)=-G_{\rm EP}\epsilon/\Gamma^2$. For coherent light, its QFI growth rate is $f_{\rm opt}=4\eta\kappa_c|\partial_\epsilon\langle a\rangle_\epsilon(\infty)|^2$ (End Matter).

Now we compare this with a non-EP mechanical oscillator. Consider a vacuum oscillator $c$ with Hamiltonian $\epsilon(c+c^\dagger)$ and jump operator, i.e., output coupling $L = \sqrt{2\Gamma}c$, whose emission due to mechanical quantum jumps is collected ideally. Its signal also has variance one and its correlation decays as $e^{-\Gamma t}$, with the QFI rate $f_{\rm r}=8/\Gamma$ for this signal. Defining the relative optical information rate as $\mathcal{E}_{\rm opt}=f_{\rm opt}/f_{\rm r}=\eta u^2(1+u)$, the direct coherent-output calculation gives $T_{\rm disc}^{\rm opt}(p,\epsilon)/T_{\rm disc}^{\rm r}(p,\epsilon)\to\mathcal{E}_{\rm opt}^{-1}$
as $\epsilon\to0$, for any fixed target detection probability $p\in(1/2,1)$, thereby reproducing the observation-time relation in Eq. (\ref{gain-detection}) for the collected cavity output. Based on demonstrated MHz-scale optical linewidths together with substantially narrower mechanical linewidths \cite{Schliesser_2008,Weis_2010,Verhagen_2012}, we use the illustrative parameters $\kappa_c/(2\pi)=6.0~\mathrm{MHz}$ and
$\gamma_m/(2\pi)=3.6~\mathrm{kHz}$, and assume ideal collection of the cavity emission, $\eta=1$. These parameters give $\Gamma/(2\pi)=1.5009~\mathrm{MHz}$,
$u\simeq0.9988$, and $\mathcal{E}_{\rm opt}\simeq1.9940$, so that $\mathcal{E}_{\rm opt}^{-1}\simeq0.5015$. Thus the observed cavity output reaches the same prescribed detection reliability with approximately $49.85\%$ less observation time than the reference output in the weak-force limit, as illustrated in Fig. \ref{fig2}(b); this is consistent with the maximum factor-of-two advantage from EP2. For an illustrative $\epsilon=10^3~\mathrm{s}^{-1}$ satisfying $\epsilon\ll\Gamma,G_{\rm EP}$, a $90\%$ detection reliability in our model requires approximately $2.42~\mathrm{s}$ at the EP, compared with $4.82~\mathrm{s}$ for the non-EP reference. More generally, an EP$m$ can generate an $m$-fold advantage if the selected perturbation to sense saturates the bound in Eq. (\ref{gain-detection}). A recent experimental architecture involving a nanomechanical cavity shows promising results for reaching EPs with $m=7$ or more \cite{Wu_2025}, so order-of-magnitude advantages may become accessible in the near future.

\section{Conclusions}
We have established fundamental detection limits on reliable quantum discrimination at Liouvillian EPs in finite-dimensional gapped Lindblad systems with unique stationary states, showing that enhanced spectral sensitivity does not improve how the required observation time grows as the perturbation becomes smaller. However, by selecting suitable perturbations, we show that an EP sensor can achieve the same detection reliability as the specified non-EP reference in less observation time, as particularly illustrated through the example of optomechanical force sensing. Our general results, applying across sensor designs therefore open up new directions for designing quantum sensors around the perturbations they must detect, with promising experimental platforms including superconducting qubits \cite{Chen_2022}, trapped ions \cite{Wu_2026}, and optomechanical force sensors \cite{Mason_2019}.\\

\noindent
\textbf{Acknowledgements:} The authors acknowledge the use of GPT-6 Sol to cross-check the derivations in the Appendices below. The content has been critically reviewed by the authors, who take full responsibility for the validity of the results. A.G. and M.B. thank the Air Force Office of Scientific Research (FA9550-23-1-0259) for support.\\

\noindent
\textbf{Data availability:}
No data were created or analyzed in this work.

\appendix

\section{Overlap evolution and detection limits}

For the finite-dimensional models, we assume analytic Hamiltonians and couplings, a simple stationary eigenvalue 0, and the fixed relaxation gap $\Delta>0$. For the common pure-state preparation $\rho_{\rm in}$, we define $X_{\delta,T}=\operatorname{Tr}_{\rm outputs}
\left[|\Psi_\delta^{(T)}\rangle\langle\Psi_0^{(T)}|\right]$, whose trace is $A_T$ and $X_{\delta,0}=\rho_{\rm in}$. This obeys the linear dynamics $\dot{X}_{\delta,T}=M_\delta(X_{\delta,T})$, where
\cite{Gammelmark_2014,Molmer_2015,Kiilerich_2018}
\begin{equation}
\begin{aligned}
M_\delta(X)
&=-i(H_\delta X-XH_0)+\sum_kL_{k,\delta}XL_{k,0}^\dagger\\
&-\frac12\sum_k\left(
L_{k,\delta}^\dagger L_{k,\delta}X+XL_{k,0}^\dagger L_{k,0}\right), 
\end{aligned}
\label{em:overlap-generator}
\end{equation}
so at $\delta=0$, one has $M_0=\mathcal{L}_0$. Its stationary eigenvalue $\chi(0)=0$ remains simple and isolated even when the decay modes coalesce. It therefore continues analytically to $\chi(\delta)$ \cite{Kato_1995}. For sufficiently small $|\delta|$, $\chi(\delta)$ remains close to zero while all other contributions decay on finite
relaxation timescales. The long-time overlap is therefore dominated by this single eigenvalue:
\begin{equation}
A_T(\delta)=\alpha(\delta)e^{T\chi(\delta)}
+O(e^{- T\Delta/2}),\qquad \alpha(0)=1,
\label{em:overlap-branch}
\end{equation}
where $\alpha(\delta)$ is analytic near $\delta = 0$. The remainder and its first two parameter derivatives decay exponentially uniformly in this neighborhood. So the pure-state QFI is given by \cite{Braunstein_1994}
\begin{equation}
\mathcal{I}_Q(T,0)=-4\left.\partial_\delta^2\log|A_T|\right|_0
=Tf_Q+O(1).
\label{em:information-rate}
\end{equation}
Since $|A_T|\leq1$, the dominant eigenvalue must satisfy $\operatorname{Re}[\chi(\delta)]\leq0$. Its value at zero is therefore a maximum, so $\operatorname{Re}[\chi'(0)]=0$, meaning that from Eq. (\ref{em:information-rate}) we identify
\begin{equation}
f_Q=-4\operatorname{Re}[\chi''(0)]\geq0, \quad \quad  \operatorname{Re}[\chi(\delta)]=-\frac{f_Q}{8}\delta^2+O(\delta^3).  
\label{fQ-expression}
\end{equation}
We now take $\delta\to0$ and $T\to\infty$ together,
with $\delta^2T$ staying bounded. In this limit, the $\alpha(\delta)$ approaches one, the relaxation transient vanishes, and the neglected exponent satisfies $|\delta|^3T=|\delta|(\delta^2T)\to0$.
Consequently, $|A_T|^2=e^{-f_Q\delta^2T/4}+o(1)$,
which gives Eq. (\ref{local-correct}) through substitution into Eq. (\ref{correct-overlap}). For fixed $T$, $\delta \rightarrow 0$ instead gives $P^\star\to1/2$.

\section{Stationary correlations and the EP balance}

For $H_\delta=H_0+\delta V$ and fixed couplings, we have from Eq. (\ref{em:overlap-generator}), the expression $M_\delta(X)=\mathcal{L}_0(X)-i\delta VX$.
Let $M_\delta(r_\delta)=\chi(\delta)r_\delta$, with
$\operatorname{Tr}[r_\delta]=1$ and $r_0=\pi$.
Taking the trace of
$M_\delta(r_\delta)=\chi(\delta)r_\delta$ gives
$\chi(\delta)=-i\delta\operatorname{Tr}(Vr_\delta)$.
Differentiating this identity with respect to $\delta$ at $\delta=0$ gives $\chi'(0)=-i\operatorname{Tr}(V\pi)=0$. Differentiating the eigenvalue equation $M_\delta(r_\delta)=\chi(\delta)r_\delta$ with respect to $\delta$ at zero then gives $\mathcal{L}_0(r'_0)-iV\pi=\chi'(0)\pi$, hence $\mathcal{L}_0(r'_0)=iV\pi$. The normalization also gives $\operatorname{Tr}[r'_0]=0$.

Because $V\pi$ is traceless, it has no stationary
component. The relaxation gap therefore ensures that
$e^{t\mathcal{L}_0}(V\pi)$ decays exponentially, making its time integral convergent. Integrating
$\partial_t[e^{t\mathcal{L}_0}(V\pi)]
=\mathcal{L}_0 e^{t\mathcal{L}_0}(V\pi)$ gives
$\mathcal{L}_0\int_0^\infty e^{t\mathcal{L}_0}(V\pi) dt=-V\pi$. This determines $r'_0$ uniquely among traceless operators. Differentiating $\chi(\delta)=-i\delta\operatorname{Tr}(Vr_\delta)$ twice also gives $\chi''(0)=-2i\operatorname{Tr}(Vr'_0)$. Consequently, we can write 
\begin{equation}
\begin{aligned}
r'_0&=-i\int_0^\infty e^{t\mathcal{L}_0}(V\pi) dt,\\
\chi''(0)&=-2\int_0^\infty
\operatorname{Tr}\!\left[V e^{t\mathcal{L}_0}(V\pi)\right]dt .
\end{aligned}
\label{em:hamiltonian-correlation}
\end{equation}
Using
$\operatorname{Tr}[V e^{t\mathcal{L}_0}(V\pi)]
=\operatorname{Tr}[\pi V(t)V]$
and taking the real part recovers
Eq.~(\ref{correlation-rate}) through
$f_Q=-4\operatorname{Re}[\chi''(0)]$
[Eq. (\ref{fQ-expression})]. It may be worth noting that no assumption that $\pi$ has full rank is needed.

Now for EP$m$, on the $m$-dimensional space of the Jordan-transient coefficients, we decompose $V$ in a Hermitian basis as (with $\sigma >0$)
\begin{equation}
\begin{aligned}
V=\sigma\sum_{\mu=1}^m v_\mu V_\mu,\qquad
\sum_{\mu=1}^m v_\mu^2&=1,\quad v_\mu\in\mathbb R,\\
\frac12\operatorname{Tr}[\pi\{V_\mu,V_\nu\}]&=\delta_{\mu\nu}.
\end{aligned}
\label{em:EP-normalization}
\end{equation}
For fixed $\sigma>0$, every such $V$ has variance
$\sigma^2$. Uniform averaging over these coefficient
vectors gives $\overline{v_\mu v_\nu}=\delta_{\mu\nu}/m$. Next we define a matrix  $B$ by $\mathcal{L}_0^\dagger V_\nu=\sum_\mu B_{\mu\nu}V_\mu$, so the coefficients evolve as $e^{tB}v$. At the EP, $B=-\gamma_{\rm EP}I+N$, with
$N^m=0$ and $N^{m-1}\neq0$. Consequently, one has
\begin{equation}
\begin{aligned}
C_V(t)&=\sigma^2v^{\mathsf T}e^{tB}v,\\
\overline{C_V(t)}
&=\frac{\sigma^2}{m}\operatorname{tr}(e^{tB})
=\sigma^2e^{-\gamma_{\rm EP}t}.
\end{aligned}
\label{em:EP-rate-matrix}
\end{equation}
Here $\operatorname{tr}$ denotes the ordinary trace of an $m\times m$ matrix. The last equality follows by expanding $e^{tB}=e^{-\gamma_{\rm EP}t}
\sum_{k=0}^{m-1}t^kN^k/k!$, where every positive power of the nilpotent matrix $N$ has zero trace. Thus all the polynomial contributions cancel before integration in Eq. (\ref{correlation-rate}), giving $\overline{f_Q}=8\sigma^2/\gamma_{\rm EP}=f_{\rm r}$. The rates of the $m$ number of $\sigma V_\mu$ therefore sum to $mf_{\rm r}$. Any signal of variance $\sigma^2$ can be included in such an orthogonal basis. Since all remaining
rates are nonnegative, its rate cannot exceed
the full sum. Hence $0\leq f_Q(V)\leq mf_{\rm r}$,
proving Eq. (\ref{EP-balance}).

\section{Optomechanical force detection}

The cavity and mechanical fluctuation modes start in their joint vacuum $\rho_\epsilon(0)=\rho_0(0)=|0,0\rangle\langle0,0|$.
Their mean amplitudes $\boldsymbol z_\epsilon(t)
=(\langle a\rangle_\epsilon(t),\langle b\rangle_\epsilon(t))^{\mathsf T}$ obey
$\dot{\boldsymbol z}_\epsilon=K\boldsymbol z_\epsilon-i\epsilon(0,1)^{\mathsf T}$,
with $\boldsymbol z_\epsilon(0)=0$ and
\begin{equation}
K=\begin{pmatrix}
-\kappa_c/2&-iG\\
-iG&-\gamma_m/2
\end{pmatrix}, \quad \quad
K_{\rm EP}=-\Gamma I+N,
\label{em:opto-EP}
\end{equation}
where $G_{\rm EP}=(\kappa_c-\gamma_m)/4$, $\Gamma=(\kappa_c+\gamma_m)/4$, and $N^2=0$ with $N\neq0$.
Shifting both the modes by their constant long-time amplitudes removes the force from the generator, preserving its decay eigenvalues and the EP.

We define the two perturbation operators $V_a=i(a^\dagger-a)$ and $V_b=b+b^\dagger$. For $\pi=|0,0\rangle\langle0,0|$, we have
$\sigma_{V_a}^2=\sigma_{V_b}^2=1$ and
$\operatorname{Tr}[\pi\{V_a,V_b\}]/2=0$. At the EP, their $\epsilon=0$ evolution is
\begin{equation}
\begin{aligned}
\mathcal{L}_0^\dagger (V_a)
&=-(\kappa_c/2)V_a-G_{\rm EP}V_b,\\
\mathcal{L}_0^\dagger (V_b)
&=G_{\rm EP}V_a-(\gamma_m/2)V_b,
\end{aligned}
\end{equation}
which give the correlations $C_{V_a}(t)=(1-G_{\rm EP}t)e^{-\Gamma t}$ and $C_{V_b}(t)=(1+G_{\rm EP}t)e^{-\Gamma t}$.

We perturb the sensor selectively only by the mechanical force $\epsilon V_b$, for which the mode operators obey
\begin{equation}
\begin{aligned}
\mathcal{L}_\epsilon^\dagger(a)
&=-\frac{\kappa_c}{2}a-iGb,\\
\mathcal{L}_\epsilon^\dagger(b)
&=-\frac{\gamma_m}{2}b-iGa-i\epsilon I.
\end{aligned}
\end{equation}
Setting the time derivatives of their mean amplitudes to zero gives $(\kappa_c/2)\langle a\rangle_\epsilon(\infty)+iG\langle b\rangle_\epsilon(\infty)=0$ and $iG\langle a\rangle_\epsilon(\infty)+(\gamma_m/2)\langle b\rangle_\epsilon(\infty)=-i\epsilon$. Thus $\langle b\rangle_\epsilon(\infty)=i\kappa_c\langle a\rangle_\epsilon(\infty)/(2G)$ and $(\kappa_c\gamma_m/4+G^2)\langle a\rangle_\epsilon(\infty)=-G\epsilon$. At the EP, we obtain 
\begin{equation}
\langle a\rangle_\epsilon(\infty)
=-\frac{G_{\rm EP}\epsilon}{\Gamma^2}, \quad \quad  \langle b\rangle_\epsilon(\infty)
=-\frac{i\kappa_c\epsilon}{2\Gamma^2},  
\label{steadystate-mech-drive}
\end{equation}
after using $\kappa_c\gamma_m/4+G_{\rm EP}^2=\Gamma^2$. At the EP, the matrix $K_{\rm EP}$ acts on the mean-amplitude vector $\boldsymbol z_\epsilon(t)$. Starting from zero initial amplitudes, its driven evolution gives
$\boldsymbol z_\epsilon(t)=-i\epsilon\int_0^t e^{sK_{\rm EP}}(0,1)^{\mathsf T}ds$. Using $e^{sK_{\rm EP}}=e^{-\Gamma s}(I+sN)$, the first component gives the cavity amplitude as
\begin{equation}
\begin{aligned}
\langle a\rangle_\epsilon(t)
&=-G_{\rm EP}\epsilon\int_0^t s e^{-\Gamma s} ds\\
&=-\frac{G_{\rm EP}\epsilon}{\Gamma^2}
[1-(1+\Gamma t)e^{-\Gamma t}].
\end{aligned}
\label{em:cavity-amplitude}
\end{equation}
The collected light has the amplitude $\sqrt{\eta\kappa_c}\langle a\rangle_\epsilon(t)$. The input-output relation gives the collected optical field a coherent amplitude $\beta_\epsilon(t)=\sqrt{\eta\kappa_c}
\langle a\rangle_\epsilon(t)$, since the incident field is in vacuum. For continuous-mode coherent states, the squared overlap of two output fields with amplitudes $\beta_\epsilon(t)$ and $\beta_0(t)$ is $\exp[-\int_0^T|\beta_\epsilon(t)-\beta_0(t)|^2dt]$. The squared overlap between the collected output states for $\epsilon\neq0$ and $\epsilon=0$ is therefore $e^{-Q_T}$, where $Q_T=\eta\kappa_c\int_0^T |\langle a\rangle_\epsilon(t)|^2dt$. Substituting Eq. (\ref{em:cavity-amplitude}) into this integral gives, as $\Gamma T\to\infty$, the expression
\begin{equation}
Q_T=\frac{\eta\kappa_cG_{\rm EP}^2\epsilon^2}{\Gamma^4}[T+O(\Gamma^{-1})],    
\end{equation}
with the sensor parameters fixed as the limit $\Gamma T\to\infty$ is taken. The coherent-state QFI $\mathcal{I}_{Q,\rm opt}=4Q_T/\epsilon^2$ 
then gives its growth rate as
\begin{equation}
f_{\rm opt}
=\lim_{T\to\infty}\frac{4Q_T}{\epsilon^2T}
=\frac{4\eta\kappa_cG_{\rm EP}^2}{\Gamma^4}.
\label{em:opto-overlap}
\end{equation}
Now we consider the non-EP reference oscillator
$\partial_t\langle c\rangle_\epsilon(t)
=-\Gamma\langle c\rangle_\epsilon(t)-i\epsilon$
with $\langle c\rangle_\epsilon(0)=0$ gives
$\langle c\rangle_\epsilon(t)
=-i\epsilon(1-e^{-\Gamma t})/\Gamma$.
Its ideally collected output therefore has
$Q_T^{\rm r} =2\Gamma\int_0^T|\langle c\rangle_\epsilon(t)|^2dt =(2\epsilon^2/\Gamma)[T+O(\Gamma^{-1})]$ and $f_{\rm r}=8/\Gamma$.

For either coherent output, the target probability $p$ requires the corresponding $Q$ to reach $-\log[4p(1-p)]$ from Eq. (\ref{opto-correct}). Since both the integrals increase continuously from zero, their first crossings determine the required observation times as
\begin{equation}
\begin{aligned}
T_{\rm disc}^{\rm opt}(p,\epsilon)
&=\frac{4}{f_{\rm opt}\epsilon^2}
\log\frac1{4p(1-p)}+O(\Gamma^{-1}),\\
T_{\rm disc}^{\rm r}(p,\epsilon)
&=\frac{4}{f_{\rm r}\epsilon^2}
\log\frac1{4p(1-p)}+O(\Gamma^{-1}).
\end{aligned}
\label{em:detection-times}
\end{equation}
For fixed $p$ and $\epsilon\to0$, the leading terms diverge and the bounded corrections become negligible. The common logarithmic factor and $\epsilon^2$ cancel in the ratio. With $u=G_{\rm EP}/\Gamma$ and $\kappa_c=2\Gamma(1+u)$, this gives
\begin{equation}
\frac{T_{\rm disc}^{\rm opt}(p,\epsilon)}
     {T_{\rm disc}^{\rm r}(p,\epsilon)}
\xrightarrow[\epsilon\to0]{}
\frac{f_{\rm r}}{f_{\rm opt}}
=[\eta u^2(1+u)]^{-1}.
\end{equation}
The collected cavity light therefore reaches the target sooner when $\eta u^2(1+u)>1$. Taking an illustrative $\epsilon=10^3~\mathrm{s}^{-1}$, which satisfies $\epsilon\ll\Gamma,G_{\rm EP}$, Eqs. (\ref{em:opto-overlap}) and (\ref{em:detection-times}) give $T_{\rm disc}^{\rm opt}\simeq2.42~\mathrm{s}$ and $T_{\rm disc}^{\rm r}\simeq4.82~\mathrm{s}$ at $p=0.9$, using $\eta=1$ and the parameters of the main text.

Finally, let us remark that the QFI in the two modes approaches a constant as their amplitudes settle for long times. If both the mechanical and optical signals are made available, the QFI rates add and give
\begin{equation}
f_Q(V)=4\kappa_c
|\partial_\epsilon\langle a\rangle_\epsilon(\infty)|^2+4\gamma_m
|\partial_\epsilon\langle b\rangle_\epsilon(\infty)|^2.
\label{em:opto-full-rate}
\end{equation}
For the mechanical perturbation, substituting Eq. (\ref{steadystate-mech-drive}) gives $f_Q(V_b)=8(1+u)/\Gamma$. If instead we considered the optical perturbation $\epsilon V_a=i\epsilon(a^\dagger-a)$, the stationary equations instead give $\langle a\rangle_\epsilon(\infty)=\gamma_m\epsilon/(2\Gamma^2)$ and $\langle b\rangle_\epsilon(\infty)
=-iG_{\rm EP}\epsilon/\Gamma^2$. Substituting these amplitudes into Eq. (\ref{em:opto-full-rate}) gives $f_Q(V_a)=8(1-u)/\Gamma$. This directly shows that while the mechanical perturbation increases the QFI rate, the optical perturbation reduces it, with their sum equaling twice the QFI rate of the reference non-EP oscillator. So the advantage is perturbation selective.


\begin{thebibliography}{99}

\bibitem{Kato_1995} 
T. Kato, \textit{Perturbation theory for linear operators}, 2nd ed., Springer (1995).

\bibitem{Berry_2004}
M. V. Berry, {\it Physics of nonhermitian degeneracies}, Czech. J. Phys. \textbf{54}, 1039 (2004).

\bibitem{Heiss_2012} 
W. D. Heiss, {\it The physics of exceptional points}, J. Phys. A: Math. Theor. \textbf{45}, 444016 (2012).

\bibitem{Miri_2019} 
M.-A. Miri and A. Al\`u, {\it Exceptional points in optics and photonics}, Science \textbf{363}, eaar7709 (2019).

\bibitem{Ashida_2020} 
Y. Ashida, Z. Gong, and M. Ueda, {\it Non-Hermitian physics}, Adv. Phys. \textbf{69}, 249 (2020)

\bibitem{Bergholtz_2021} 
E. J. Bergholtz, J. C. Budich, and F. K. Kunst, {\it Exceptional topology of non-Hermitian systems}, Rev. Mod. Phys. \textbf{93}, 015005 (2021).

\bibitem{Wiersig_2022a}
J. Wiersig, {\it Response strengths of open systems at exceptional points}, Phys. Rev. Research \textbf{4}, 023121 (2022).

\bibitem{Wiersig_2022b}
J. Wiersig, {\it Distance between exceptional points and diabolic points and its implication for the response strength of non-Hermitian systems}, Phys. Rev. Research \textbf{4}, 033179 (2022).

\bibitem{Wiersig_2014}
J. Wiersig, {\it Enhancing the sensitivity of frequency and energy splitting detection by using exceptional points: Application to microcavity sensors for single-particle detection}, Phys. Rev. Lett. \textbf{112}, 203901 (2014).

\bibitem{Wiersig_2016}
J. Wiersig, {\it Sensors operating at exceptional points: General theory}, Phys. Rev. A \textbf{93}, 033809 (2016).

\bibitem{Chen_2017}
W. Chen, \c{S}. K. \"Ozdemir, G. Zhao, J. Wiersig, and L. Yang, {\it Exceptional points enhance sensing in an optical microcavity}, Nature \textbf{548}, 192 (2017).

\bibitem{Hodaei_2017}
H. Hodaei, A. U. Hassan, S. Wittek, H. Garcia-Gracia, R. El-Ganainy, D. N. Christodoulides, and M. Khajavikhan, {\it Enhanced sensitivity at higher-order exceptional points}, Nature \textbf{548}, 187 (2017).

\bibitem{Chen_2018}
P.-Y. Chen, M. Sakhdari, M. Hajizadegan, Q. Cui, M. M.-C. Cheng, R. El-Ganainy, and A. Al\`u, {\it Generalized parity-time symmetry condition for enhanced sensor telemetry}, Nat. Electron. \textbf{1}, 297 (2018).

\bibitem{Lai_2019}
Y.-H. Lai, Y.-K. Lu, M.-G. Suh, Z. Yuan, and K. Vahala, {\it Observation of the exceptional-point-enhanced Sagnac effect}, Nature \textbf{576}, 65 (2019).

\bibitem{Hokmabadi_2019}
M. P. Hokmabadi, A. Schumer, D. N. Christodoulides, and M. Khajavikhan, {\it Non-Hermitian ring laser gyroscopes with enhanced Sagnac sensitivity}, Nature \textbf{576}, 70 (2019). 

\bibitem{Park_2020}
J.-H. Park, A. Ndao, W. Cai, L. Hsu, A. Kodigala, T. Lepetit, Y.-H. Lo, and B. Kant\'e, {\it Symmetry-breaking-induced plasmonic exceptional points and nanoscale sensing}, Nat. Phys. \textbf{16}, 462 (2020). 

\bibitem{Wiersig_2020}
J. Wiersig, {\it Review of exceptional point-based sensors}, Photon. Res. \textbf{8}, 1457 (2020).

\bibitem{Langbein_2018} 
W. Langbein, {\it No exceptional precision of exceptional-point sensors}, Phys. Rev. A \textbf{98}, 023805 (2018).

\bibitem{Lau_2018} H.-K. Lau and A. A. Clerk, {\it Fundamental limits and non-reciprocal approaches in non-Hermitian quantum sensing}, Nat. Commun. \textbf{9}, 4320 (2018).

\bibitem{Zhang_2019}
M. Zhang, W. Sweeney, C. W. Hsu, L. Yang, A. D. Stone, and L. Jiang, {\it Quantum noise theory of exceptional point amplifying sensors}, Phys. Rev. Lett. \textbf{123}, 180501 (2019).

\bibitem{Chen_2019}
C. Chen, L. Jin, and R.-B. Liu, {\it Sensitivity of parameter estimation near the exceptional point of a non-Hermitian system}, New J. Phys. \textbf{21}, 083002 (2019).

\bibitem{Ding_2023}
W. Ding, X. Wang, and S. Chen, {\it Fundamental sensitivity limits for non-Hermitian quantum sensors}, Phys. Rev. Lett. \textbf{131}, 160801 (2023).

\bibitem{Loughlin_2024} H. Loughlin and V. Sudhir, {\it Exceptional-point sensors offer no fundamental signal-to-noise ratio enhancement}, Phys. Rev. Lett. \textbf{132}, 243601 (2024).

\bibitem{Peng_2024}
J.-X. Peng, B. Zhu, W. Zhang, and K. Zhang, {\it Dissipative quantum Fisher information for a general Liouvillian parametrized process}, Phys. Rev. A \textbf{109}, 012432 (2024).

\bibitem{Wiersig_2026}
J. Wiersig and S. Rotter, {\it Fundamental limits of non-Hermitian sensing from quantum Fisher information}, Rep. Prog. Phys. \textbf{89}, 067501 (2026).

\bibitem{Wang_2026}
C. Wang, D. Hu, S. Zorzetti, A. Grassellino, A. Romanenko, and Z. Zhang, {\it Squeezing-enhanced sensing at an exceptional point}, Phys. Rev. Lett. \textbf{136}, 250801 (2026).

\bibitem{Holevo_1973}
A. S. Holevo, {\it Statistical decision theory for quantum systems}, J. Multivariate Anal. \textbf{3}, 337 (1973).

\bibitem{Helstrom_1976} C. W. Helstrom, {\it Quantum detection and estimation theory}, Academic Press (1976).

\bibitem{Barnett_2009}
S. M. Barnett and S. Croke,
{\it Quantum state discrimination},
Adv. Opt. Photon. \textbf{1}, 238 (2009).

\bibitem{Watrous_2018}
J. Watrous, {\it The theory of quantum information}, Cambridge University Press (2018).

\bibitem{Molmer_2015}
K. M{\o}lmer, {\it Hypothesis testing with open quantum systems}, Phys. Rev. Lett. \textbf{114}, 040401 (2015).

\bibitem{Kiilerich_2018}
A.~H. Kiilerich and K. M{\o}lmer, {\it Multistate and multihypothesis discrimination with open quantum systems}, Phys. Rev. A \textbf{97}, 052113 (2018).

\bibitem{Chen_QSD_2022}
D.-X. Chen, Y. Zhang, J.-L. Zhao, Q.-C. Wu, Y.-L. Fang, C.-P. Yang, and F. Nori, {\it Quantum state discrimination in a $\mathcal{PT}$-symmetric system}, Phys. Rev. A \textbf{106}, 022438 (2022).

\bibitem{Bouchet_2021}
D. Bouchet, L. M. Rachbauer, S. Rotter, A. P. Mosk,
and E. Bossy, {\it Optimal control of coherent light scattering for binary decision problems}, Phys. Rev. Lett. \textbf{127}, 253902 (2021).

\bibitem{Gorini_1976}
V. Gorini, A. Kossakowski, and E. C. G. Sudarshan, {\it Completely positive dynamical semigroups of $N$-level systems}, J. Math. Phys. \textbf{17}, 821 (1976).

\bibitem{Lindblad_1976}
G. Lindblad, {\it On the generators of quantum dynamical semigroups}, Commun. Math. Phys. \textbf{48}, 119 (1976).

\bibitem{Chen_2022}
W. Chen, M. Abbasi, B. Ha, S. Erdamar, Y. N. Joglekar, and K. W. Murch, {\it Decoherence-induced exceptional points in a dissipative superconducting qubit}, Phys. Rev. Lett. \textbf{128}, 110402 (2022).

\bibitem{Hu_2018}
L. Hu, X. Mu, W. Cai, Y. Ma, Y. Xu, H. Wang,
Y. P. Song, C.-L. Zou, and L. Sun, {\it Experimental repetitive quantum channel simulation}, Sci. Bull. \textbf{63}, 1551 (2018).

\bibitem{Wu_2026}
Z.-Z. Wu, P.-D. Li, T.-H. Cui, J.-W. Wang,
Y.-Z. Dong, S.-Q. Dai, J. Li, Y.-Q. Wei,
Q. Yuan, X.-M. Cai, L. Chen, J.-Q. Zhang,
H. Jing, and M. Feng, {\it Experimental witness of quantum jump induced high-order Liouvillian exceptional points}, Nat. Commun. \textbf{17}, 1923 (2026).

\bibitem{Abo_2024}
S. Abo, P. Tulewicz, K. Bartkiewicz, \c{S}. K. \"Ozdemir, and A. Miranowicz, {\it Experimental Liouvillian exceptional points in a quantum system without Hamiltonian singularities}, New J. Phys. \textbf{26}, 123032 (2024).

\bibitem{Minganti_2019} 
F. Minganti, A. Miranowicz, R. W. Chhajlany, and F. Nori, {\it Quantum exceptional points of non-Hermitian Hamiltonians and Liouvillians: The effects of quantum jumps}, Phys. Rev. A \textbf{100}, 062131 (2019).

\bibitem{Minganti_2020}
F. Minganti, A. Miranowicz, R. W. Chhajlany, I. I. Arkhipov, and F. Nori, {\it Hybrid-Liouvillian formalism connecting exceptional points of non-Hermitian Hamiltonians and Liouvillians via postselection of quantum trajectories}, Phys. Rev. A \textbf{101}, 062112 (2020).

\bibitem{Khandelwal_2021}
S. Khandelwal, N. Brunner, and G. Haack, {\it Signatures of Liouvillian exceptional points in a quantum thermal machine}, PRX Quantum \textbf{2}, 040346 (2021).

\bibitem{Catana_2015}
C. Catana, L. Bouten, and M. Gu\c{t}\u{a}, {\it Fisher informations and local asymptotic normality for continuous-time quantum Markov processes}, J. Phys. A: Math. Theor. \textbf{48}, 365301 (2015).

\bibitem{Guta_2017}
M. Gu\c{t}\u{a} and J. Kiukas,
{\it Information geometry and local asymptotic normality for multi-parameter estimation of quantum Markov dynamics}, J. Math. Phys. \textbf{58}, 052201 (2017).

\bibitem{Tsang_2012}
M. Tsang and R. Nair, {\it Fundamental quantum limits to waveform detection}, Phys. Rev. A \textbf{86}, 042115 (2012).

\bibitem{Mason_2019}
D. Mason, J. Chen, M. Rossi, Y. Tsaturyan,
and A. Schliesser, {\it Continuous force and displacement measurement below the standard quantum limit}, Nat. Phys. \textbf{15}, 745 (2019).

\bibitem{Aspelmeyer_2014} M. Aspelmeyer, T. J. Kippenberg, and F. Marquardt, {\it Cavity optomechanics}, Rev. Mod. Phys. \textbf{86}, 1391 (2014).

\bibitem{Ghosh_2026}
A. Ghosh and M. Bhattacharya, {\it Quantum jumps in open cavity optomechanics and Liouvillian versus Hamiltonian exceptional points}, Phys. Rev. A \textbf{113}, 053530 (2026).

\bibitem{Braunstein_1994} S. L. Braunstein and C. M. Caves, {\it Statistical distance and the geometry of quantum states}, Phys. Rev. Lett. \textbf{72}, 3439 (1994).

\bibitem{Gammelmark_2014}
S. Gammelmark and K. M{\o}lmer, {\it Fisher information and the quantum Cram\'er-Rao sensitivity limit of continuous measurements}, Phys. Rev. Lett. \textbf{112}, 170401 (2014).

\bibitem{Yokomizo_2026}
K. Yokomizo, A. A. Clerk, and Y. Ashida,
{\it Asymptotic scaling of precision limits in continuous Gaussian quantum metrology},
arXiv:2601.13554.

\bibitem{Horodynski_2021}
M. Horodynski, D. Bouchet, M. K\"uhmayer, and S. Rotter, {\it Invariance property of the Fisher information in scattering media}, Phys. Rev. Lett. \textbf{127}, 233201 (2021).

\bibitem{Satzinger_2018}
K. J. Satzinger, Y. P. Zhong, H.-S. Chang, G. A. Peairs, A. Bienfait, M.-H. Chou, A. Y. Cleland, C. R. Conner, \'E. Dumur, J. Grebel, I. Gutierrez, B. H. November, R. G. Povey, S. J. Whiteley, D. D. Awschalom, D. I. Schuster, and A. N. Cleland,
{\it Quantum control of surface acoustic-wave phonons}, Nature \textbf{563}, 661 (2018).

\bibitem{Omahen_2026}
A. Omahen, S. Storz, M. Bild, D. Scheiwiller,
M. Fadel, and Y. Chu, {\it Ultracold mechanical quantum sensor for tests of new physics}, Phys. Rev. Lett. \textbf{136}, 180802 (2026).

\bibitem{Schliesser_2008}
A. Schliesser, R. Rivi\`ere, G. Anetsberger, O. Arcizet, and T. J. Kippenberg, {\it Resolved-sideband cooling of a micromechanical oscillator}, Nat. Phys. \textbf{4}, 415 (2008).

\bibitem{Weis_2010}
S. Weis, R. Rivi\`ere, S. Del\'eglise, E. Gavartin,
O. Arcizet, A. Schliesser, and T. J. Kippenberg, {\it Optomechanically induced transparency},
Science \textbf{330}, 1520 (2010).

\bibitem{Verhagen_2012}
E. Verhagen, S. Del\'eglise, S. Weis, A. Schliesser,
and T. J. Kippenberg, {\it Quantum-coherent coupling of a mechanical oscillator to an optical cavity mode}, Nature \textbf{482}, 63 (2012).

\bibitem{Wu_2025}
N. Wu, K. Cui, Z. Chen, C. Wang, X. Feng, F. Liu, W. Zhang, H. Sun, Y. Li, and Y. Huang, {\it Arbitrary-order exceptional points in a nanomechanical cavity}, arXiv:2512.11679.


\end{thebibliography}
\end{document}